# Should the Endless Frontier of Federal Science be Expanded?

David Baltimore[a], Robert Conn[b], William Press[c], Thomas Rosenbaum[d], David Spergel[e], Shirley Tilghman[f], and Harold Varmus[g]

February 28, 2021

**Abstract**

Scientific research in the United States could receive a large increase in federal funding—up to $100 billion over five years—if proposed legislation entitled the *Endless Frontiers Act*[1] becomes law. This bipartisan and bicameral bill, introduced in May 2020 by Senators Chuck Schumer (D-NY) and Todd Young (R-IN) and Congressmen Ro Khanna (D-CA) and Mike Gallagher (R-WI), is intended to expand the funding of the physical sciences, engineering, and technology at the National Science Foundation (NSF) and to create a new Technology Directorate focused on use-inspired research. In addition to provisions to protect the current NSF's current missions, a minimum of 15% of newly appropriated funds would be used to enhance NSF's basic science portfolio.

The *Endless Frontier Act* offers a rare opportunity to enhance the breadth and financial support of the American research enterprise. In this essay, we consider the benefits and the liabilities of the proposed legislation and recommend changes that would further strengthen it.

**i. Background and Context**

For the past 75 years, American science has prospered under the influence of a blueprint laid out by Vannevar Bush in his report *Science The Endless Frontier*[2]. Bush, who had led the Office of Scientific Research and Development during World War Two, wrote that now-classic text in response to a 1944 request from President Franklin D. Roosevelt for a plan that would allow the U.S. to benefit from scientific work as much in peacetime as it had in wartime.

The essential feature of Bush's vision was to use federal funds, largely at the nation's academic institutions, to support unfettered basic research and to train scientists in many fields. He then expected the commercial sector to make use of the discoveries and well-trained scientists generated through those activities to invent the practical goods that would fuel the economy, enhance health, strengthen the national defense, and improve the quality of life.

For the most part, Bush's formula has worked well, making the strong support of scientific work an accepted domain of federal action, even in normal times. Although Bush's initial proposal to house all federally funded science under the roof of a single agency, a National Research Foundation, was displaced by a more complex array of funding agencies, the report's fundamental features led to vast increases in the federal funding of science through the National Science Foundation and other agencies, helped to propel the U.S. to dominance in many fields of science, and are the pillars of science policy today.

The world and the conduct of science have, of course, changed over the decades, inviting efforts to revise and strengthen the ways in which the federal government supports science. Two immediate crises -- one we are struggling to control (the Covid-19 pandemic), another with no apparent end in sight (climate change) -- have highlighted the importance of science for the welfare of the nation and even the survival of our species.  More generally, the future competitiveness and vitality of the economy and the effectiveness of our defense and intelligence systems will depend to a large extent upon scientific advances.  While the U.S. still retains international prestige in the sciences, other countries, especially in Asia, have expanded their capacities, competing effectively with us in multiple domains, targeting investments to create jobs in emerging industries, and generating fears that we are losing our competitive edge.

At the same time, the conduct of science has itself become more complicated, accompanied by an increasing recognition of the nuances required for success: the convergence of multiple disciplines to study difficult problems; the imperative of generating new and expensive technologies; the enlarging role of powerful computational methods in all fields; and the long-overdue demands to achieve racial, ethnic, and gender diversity throughout the scientific workforce.

From these perspectives and others, Vannevar Bush's paradigm warrants reappraisal and revision.  Efforts to do so are justified on several grounds: to confront immediate and enormous threats that only science can reverse; to adapt the conduct of science to new conditions and competitors; and to speed the progress of science while retaining the qualities that have made the American version of the scientific enterprise pre-eminent.  The letter[3] that President Joe Biden sent last month to his science advisor, Eric Lander, reflects this new reality. President Biden is once again asking, as FDR did seventy-five years ago, how science and technology can best be used to serve the nation.

**II. How the Bill would Expand the Vision and the Funding**

The *Endless Frontier Act*[1] would go a long way toward responding to the nation's needs for an expanded and more sophisticated conception of the federal role in science.  The bill recognizes the need for greater federal investment in a specific part of the scientific ecosystem that has been under-appreciated and under-financed in the dominant model---namely, the development of complex technologies, those that drive basic discoveries in many fields and accelerate the practical applications of science.

The bill would establish within the National Science Foundation (NSF) a new Technology Directorate, alongside its several existing science and engineering directorates, to increase investment in basic sciences that are motivated by specific problems and needs.  (This is sometimes referred to as use-inspired basic research[4].) The new directorate would be authorized to receive a very large budget, as much as $100 billion over its first five years, more than the NSF's other directorates combined and a substantial increase over its FY2021 budget of approximately $8.5 billion.

To meet the needs for skilled personnel in an enlarged enterprise, the NSF would also be directed to expand its role in the training of new scientists and engineers. To diversify the locations where such work is done, the Commerce Department would be charged to create up to a dozen regional technology centers around the country, with funds that could amount to as much as $20 billion over ten years to pursue a flexible list of broad national needs. In these several ways, federal support for the scientific enterprise would be dramatically increased by emphasizing a category not explicitly described in Vannevar Bush's original formulation: use-inspired basic science and technology.

Such ambitious plans to meet 21st century requirements will encounter justifiable resistance if they do not also include commitments to protect the basic science that the NSF and other federal funding agencies already perform so well, or if the plans undermine Vannevar Bush's edict to preserve "freedom of inquiry"[2].

### III. Assessment of the Bill and Recommendations for Consideration

Over the past several months, we have been debating among ourselves and consulting with colleagues and Congressional staff about the goals to be pursued, the measures to be taken, and the safeguards to be provided under the terms of this potentially landmark legislation. Now, with a new Congress, a new Administration, and new leadership for science in the White House, the bill is likely to be revised and reintroduced in the Congressional session that began recently. We offer several observations -- about budgets, education and training, and governance -- that we believe will make a new version of the bill even stronger than the current one.

#### *Budget*

By proposing very large sums for the new Technology Directorate, the bill is likely to raise concerns that a realigned NSF will have objectives that are at odds with its traditional strengths in fundamental science and that the new directorate will overwhelm existing ones, siphoning funds away from their activities. However, the bill also authorizes additional funding for NSF's current directorates and prevents additional activities for the new directorate if the budgets of the others are not maintained (at inflation-adjusted levels) by Congressional appropriators.

These measures are comforting, but the legislation could further strengthen the guardrails for the support of fundamental science in at least three ways:

(i) by specifying minimum increases in inflation-adjusted funding of the existing directorates that would be required to allow increases for the new directorate.

(ii) by encouraging appropriators to design the ramp-up and continuation of the Technology Directorate's budget in a manner commensurate with the long-term nature of scientific work, while avoiding the kind of fiscal instability and erosion that followed the five-year doubling of the budget for the National Institutes of Health (NIH) between 1998 and 2003; and

(iii) by adjusting plans for governance, as we suggest below.

Preservation of budgets for the conduct of unrestrained basic science and for their continued growth is an important facet of any effort to change the federally financed research enterprise. But it is not the only one. The country---and the world---have also benefited from unorthodox practices that encourage highly imaginative, risky, but sometimes remarkably fruitful science.

Some of these have occurred in the private sector (e.g., Bell Laboratories, H-P's Research Center, Xerox PARC and the research labs of pharmaceutical companies); some in the public domain (e.g., the Defense Advanced Research Project Agency, the Advanced Research Project Agency-Energy, and Director's discretionary funds at Department of Energy national laboratories); and some in the philanthropic domain (e.g., the Howard Hughes Medical Institute, the Simons Foundation's Flatiron Institutes, the twenty Kavli Institutes, the Broad Institute, and recently the Chan-Zuckerberg Initiative.)

While it would be over-reaching for legislators to draw up detailed plans for the ten major regional laboratories and their programs, the bill could specifically encourage the leaders of the new NSF technology directorate to design research practices and environments likely to nurture advances that are unanticipated and unplanned.

### *Education and Training*

In recognizing the demand for a larger federally supported workforce in the many fields of science to be supported by an expanded NSF, the proposed legislation includes increased funding for undergraduate scholarships, graduate fellowships and traineeships, and postdoctoral fellowships as the preferred mechanisms for supporting trainees. This funding model also enables greater intellectual and personal mobility.

The attention to training in STEM fields (Science, Technology, Engineering and Mathematics) provides a welcome and much needed opportunity to improve the way the nation recruits and educates its talented youth. With the country's changing demography, its history of under-representation of Black, Latinx, and indigenous people in STEM occupations, and the increasing importance being accorded to this issue as its injustices have become more apparent, the new legislation should offer new means to attract a larger and more diverse set of students into STEM fields. This is not only the positive and right thing to do, it also addresses the losses we suffer when a significant portion of our population is not able to contribute to the larger effort.

We also contend that trainees and other STEM personnel should be encouraged to gain broader experiences in multiple sectors of the research ecosystem, including industry---a kind of "tech transfer" through people---to shorten the paths between scientific discovery and the development of new technologies or between those technologies and their use to make new products or perform new science.

More experiments in educating and training scientists and engineers should also be considered, such as expansion of funded master's degree programs in STEM areas, especially those of particular interest to industry. In addition, Ph.D. candidates, even those in the most fundamental areas of research, could be encouraged to obtain exposure to research in industrial settings. Such programs could help to meet the nation's future workforce needs in engineering and in the information and biomedical sciences.

### *Governance and Organization of NSF*

To achieve the full benefits of its increased budget authority and its expanded mission, we strongly believe that the NSF must retain the unity of structure that historically has made its whole greater than the sum of its parts. Specifically, it is crucial that the NSF continue to be led unambiguously by a single, Senate-confirmed Director and to be overseen by a single National Science Board, which could be expanded to ensure that it can offer sound advice to the new Technology Directorate.

The currently proposed bill, in contrast, would place Senate-confirmed administrators---appearing to out-rank the leaders of current directorates---in charge of the new one, and it would create a new advisory board that includes Congressionally appointed members. We do not support these organization changes, as they are likely to promote a collision of cultures within the NSF at a time when the performance of the agency will depend increasingly on collaboration among its various disciplines and directorates.

The legislation also proposes to change the name of the current agency, adding the word "Technology" to create the National Science and Technology Foundation or NSTF. In the interest of extending the NSF's acclaimed history and position in science, we strongly recommend that its current, well-recognized name National Science Foundation (NSF) be preserved.

### Conclusions

The recommendations we have made here are significant, but still minor in comparison with the benefits that the proposed *Endless Frontier Act* would confer on the nation's science and technology research enterprise. We recognize that many of those normally engaged with that enterprise have been distracted by the turbulence of the past year and remain relatively uninformed about the Act's provisions, so we now urge them to give the proposed legislation their constructive attention. We have ourselves concluded that the bill offers a rare, once-in-a-generation opportunity to improve our scientific future and to extend American leadership in science and technology for the benefit of the nation and the world.

**Author Affiliations**
**a.** President, Emeritus and Distinguished Professor of Biology. California Institute of Technology
**b.** President and CEO, Retired. The Kavli Foundation. Dean and Walter Zable Professor of Applied Physics, Emeritus. Jacobs School of Engineering, University of California, San Diego
**c.** Leslie Surginer Professor of Computer Science and Integrative Biology, The University of Texas at Austin
**d.** President, Davidow Presidential Chair and Professor of Physics, California Institute of Technology
**e.** Director, Center for Computational Astrophysics. Flatiron Institute, and President-elect, The Simons Foundation
**f.** President, Emerita and Professor of Molecular Biology and Public Policy, Princeton University
**g.** Lewis Thomas University Professor, Weill-Cornell Medicine. Former Director, National Institutes of Health and National Cancer Institute